\documentclass[aps,prd,twocolumn,groupedaddress]{revtex4}

\begin{document}

\preprint{INFNCA-TH0201]}

\title{Quantum dilaton gravity as a linear dilaton conformal field theory}

\author{M.~Cadoni}
\email{mariano.cadoni@ca.infn.it}
\author{P.~Carta}
\email{paolo.carta@ca.infn.it}
\affiliation{Dipartimento di Fisica,
Universit\`a di Cagliari, and INFN sezione di Cagliari, Cittadella
Universitaria 09042 Monserrato, ITALY}

\author{S.~Mignemi}
\email{smignemi@vaxca1.unica.it}
\affiliation{Dipartimento di Matematica,
Universit\`a di Cagliari, Viale Merello 92, 09123 Cagliari, ITALY
and INFN sezione di Cagliari}

\date{\today}

\begin{abstract}
A model of matter-coupled gravity in two dimensions is quantized.  The
crucial requirement for performing the quantization is the vanishing
of the conformal anomaly, which is achieved by tuning a parameter in
the interaction potential. The spectrum of the theory is determined by
mapping the model first onto a field theory with a Liouville
interaction, then onto a linear dilaton conformal field theory. In
absence of matter fields a pure gauge theory with massless ground
state is found; otherwise it is possible to minimally couple up to 11
matter scalar fields: in this case the ground state is tachyonic and the
matter sector decouples, like the transverse oscillators in the
critical bosonic string.

\end{abstract}

\pacs{Valid PACS appear here}

\maketitle

\section{Introduction}
In recent years many controversial features of the semiclassical and quantum 
theory of gravity have been studied using models in two spacetime dimensions.
Two-dimensional(2D) theories of gravity  not only are useful toy models
but in some cases also have a direct physical meaning, since they
may be obtained from dimensional reduction of higher dimensional
gravity theories.

Basically, one can approach 2D gravity from two different points of
view: as the theory of 2D random surfaces (a bosonic string theory in
non-critical dimensions) \cite{pol,kawai} or as a Einstein-like,
Brans-Dicke theory of gravity (2D dilaton gravity) \cite{dilgra,jt}.

One crucial feature of 2D dilaton gravity is that it allows for 
interesting  gravitational structures already at the classical level.
The theory admits black hole solutions, which have been 
investigated both at the classical and semiclassical level 
\cite{kunca}.
Big progress in this direction has been achieved using the Anti-de 
Sitter(AdS)/Conformal field theory (CFT) correspondence in 
two-dimensions \cite{cami,CCKM,cacav,med}.
The use of this correspondence made it possible to exactly reproduce the  
thermodynamical entropy of 2D black holes in terms of the degeneracy 
of states of a CFT. 

It is obvious that one would like to go beyond the semiclassical
approximation, in order to see how the results of the semiclassical
approach are modified in a full 2D quantum theory of gravity.  This
program has been pursued using a variety of methods \cite{jack,CAF}
but the results have been in some sense astonishing: the quantum theory
forgets almost completely the richness of the structure of the
classical and semiclassical theory. The spectrum of the model (which
in some cases is that of a CFT) is characterized only by two quantum
numbers, which can be identified with the black hole mass and the
constant mode of the dilaton.  This feature appears even more puzzling
when compared with the results of the AdS$_{2}$/CFT$_1$ correspondence
\cite{cami}, which predicts a huge degeneracy of states accounting
for the thermodynamical entropy of the black hole.

The purpose of this paper is to go a step further in this direction.
We will quantize exactly a 2D dilaton gravity model.  We will focus
our analysis on matter-coupled dilaton gravity where the gravitational
sector is given by
\begin{equation}\label{e:jack}
S = \frac{1}{k}\int d^2x \sqrt{-g} \left[ \phi R + \lambda V(\phi)\right],
\end{equation}
and we choose an exponential dilaton potential $V(\phi)= const\cdot
e^{2\gamma\phi}$.  We will approach the problem of the quantization of
the model using nonperturbative methods.  A natural way to introduce
them is to relate the gravitational model to one of the several
exactly solved quantum field theories in $1+1$ dimensions. 
The Liouville field theory (LFT) turns out to be,
not surprisingly, the right choice.  Fixing the diffeomorphism
invariance of the action (\ref{e:jack}) and using suitable field
redefinitions, we will show that the gravity model is classically
equivalent to a LFT plus a decoupled free scalar with wrong sign
kinetic energy.

We will then follow the common quantization procedure, imposing the
positivity of quantum states energy (so, as usual in string theory, we
will have ghosts in the spectrum).  A consistent quantization can be
performed provided the quantum anomaly vanishes. For this purpose, a
simple mechanism may be used. It is well known that also at the
classical level LFT has a central charge $c_\gamma$, depending on the
parameter $\gamma$ appearing in the potential $\exp(2\gamma\phi)$. In
quantum LFT $c_\gamma$ is shifted by $\gamma$ dependent quantum
corrections \cite{thorn}. It will turn out that this shift is crucial:
fixing $\gamma$ properly, we can achieve a vanishing total anomaly and
remove the obstruction to quantization. If it were not for this shift,
any dependence on the control parameter $\gamma$ would be lost and 
a nonvanishing anomaly would be unavoidable.

In a sense our approach is similar to that used by David, Distler
and Kawai in a different context \cite{kawai}. In Ref.  \cite{kawai}
the Weyl symmetry is used to determine the coefficients of a free
action (the exponential interaction is canceled by fixing the
cosmological constant of the string).  Conversely, in this paper we
will fix the parameter $\gamma$ appearing in the dilaton potential in
order to ensure the conformal symmetry.  It turns out that also in our
case the whole theory, including matter fields, may be mapped onto a
free field theory: a linear dilaton CFT (see for instance Ref. 
\cite{polc}) with
energy-momentum tensor given by 
\begin{equation}\label{e:ldcft}
T_{\pm \pm} = -\frac{1}{\alpha'} :\partial_\pm X^\mu \partial_\pm X_\mu: +
V_\mu\partial_\pm^2 X^\mu.
\end{equation}
In our case $V_\mu$ is determined by the dilaton gravity action. 

The theory described by (\ref{e:ldcft}) is not suitable as a string theory
since the Lorentz invariance in the target space is explicitly broken
by $V_\mu$. This is not important in our context, since internal
symmetries of the fields are not relevant for us.  The quantization of
the theory then follows in a straightforward way.

The structure of the paper is the following.  In sect. II we introduce
the gravitational model we are going to investigate, we discuss the
classical equivalence to constrained Liouville field theory plus a scalar free
field  and we calculate the central charge as a
function of the parameter $\gamma$. In sect. III we briefly remind some
basic feature of LFT and we map our model onto a linear dilaton
CFT. In sect. IV the spectrum of the quantum theory is determined.
Finally in sect. V we state our conclusions.

\section{The gravitational model}
Let us consider a general dilaton gravity model with $N-2$ minimally
coupled matter fields $\varphi_i$, $i=3,\ldots,N$ ($N\ge 2$, $N<14$):
\begin{eqnarray}\label{e:grav}
S &=& \frac{1}{k}\int d^2x \sqrt{-g} \left[ \phi R +K(\phi)
 (\nabla_{\mu}\phi)(\nabla^{\mu}\phi) +\right. \nonumber \\
 &&\left.\lambda W(\phi)+ L
 (\nabla_{\mu}\varphi_i)(\nabla^{\mu}\varphi_i)\right],
\end{eqnarray}
where $W(\phi)$ (the dilaton potential), $K(\phi)$ are functions of
the dilaton and $\lambda$ and $L$ are coupling constants.  At the
classical level, whereas the matter part of the action is both
diffeomorphism and Weyl invariant, the gravitational sector looks
invariant only under diffeomorphisms.  However, it has been shown that
also the gravitational sector is classically Weyl invariant \cite{c1}.  
Using this  symmetry, the action
(\ref{e:grav}) can be transformed, by means of  a dilaton-dependent Weyl
rescaling of the metric, into the action (\ref{e:jack}), with
$W(\phi)=\exp(\int d\phi K(\phi)) V(\phi)$.  Owing to the conformal
anomaly, the extension of the Weyl symmetry at the quantum level is in
general problematic. But not in our case, since we tune to zero the
conformal anomaly. (For the dilaton gravity model  with an exponential 
potential we are considering in this paper, the quantum equivalence 
between the model (\ref{e:grav}) (with $K=const.$) and (\ref{e:jack})
can be also established showing that both can be mapped into the action
(\ref{e:liouv_2}), see below).

We fix the diffeomorphism invariance of the action (\ref{e:jack}) by
choosing the conformal gauge for the metric,
\begin{equation}\label{e:gauge}
ds^{2}= e^{2\gamma\rho(x)} dx^{+}dx^{-},
\end{equation}
where $\gamma$ is a real free parameter.  A glance at the classical
theory may be useful. For the moment we consider only the
gravitational sector, matter fields will be  reintroduced
later. The equations of motion in the conformal gauge read:
\begin{eqnarray*}
&& 8\gamma e^{-2\gamma\rho(x)}\partial_+\partial_-\rho(x)=\lambda
\frac{dV(\phi)}{d\phi} \\
&& \partial_+\partial_-\phi(x)
-\frac{\lambda}{4}e^{2\gamma\rho(x)}V(\phi)=0 \\
&& \partial_+^2\phi -2\gamma\partial_+\rho(x)\partial_+\phi(x)=0\\
&& \partial_-^2\phi -2\gamma\partial_-\rho(x)\partial_-\phi(x)=0,
\end{eqnarray*}
where the last two equations are the constraints for the theory. Two
choices for $V(\phi)$ lead to integrable models suitable for our
purposes: $V(\phi)=\phi$ and $V(\phi)=\alpha e^{2\gamma\phi}+\beta
e^{-2\gamma\phi}$ \cite{fil}. The first case was considered in Refs.
\cite{jack2,jack}. Classical solutions of the latter have been
discussed in \cite{FN}. Here we will focus on the second case.  Later
on, we will show that a quantum treatment of the model is not possible
for generic values of $\alpha$ and $\beta$.  We choose $\alpha=1$ and
$\beta=0$. Setting $k=\pi\gamma$, $\lambda=\mu/(8\gamma)$ and
performing the field redefinitions:
\begin{displaymath}
\psi = 2(\rho+\phi) \qquad \qquad \chi = 2(\rho-\phi),
\end{displaymath}
action (\ref{e:jack}) in the conformal gauge (\ref{e:gauge}) becomes:

\begin{equation}\label{e:liouv_2}S = \frac{1}{4\pi}\int dx^+dx^- \left(
\partial_+\psi\partial_-\psi
-\partial_+\chi\partial_-\chi
+\frac{\mu}{4\gamma^2}e^{\gamma\psi} \right),
\end{equation}
whereas the constraint equations are
\begin{equation}\label{e:constr_1}
T_{\pm\pm} = -\frac{1}{2}(\partial_\pm\psi)^2 +
\frac{1}{\gamma}\partial^2_\pm\psi - (-\frac{1}{2}(\partial_\pm\chi)^2
+ \frac{1}{\gamma}\partial^2_\pm\chi) =0.
\end{equation}
The action (\ref{e:liouv_2}) is invariant, up to boundary terms, under
transformations of the conformal group in two dimensions, given in
light-cone coordinates by $x^+\to w^+(x^+),\, x^- \to w^-(x^-)$. Under
these transformations the field $\chi$ transforms as a scalar, whereas
$\psi$ transforms as a Liouville field $\psi\to\psi +(1/\gamma)\ln
[(dw^+/dx^-)(dw^-/dx^-)]$.

The physical content of the 2D field theory we have obtained can be
immediately read from Eqs. (\ref{e:liouv_2}) and (\ref{e:constr_1}).
It is given by Liouville theory for the field $\psi$ and one decoupled
free scalar field $\chi$ with wrong sign kinetic energy. Both fields
have an improvement term in the stress-energy tensor. Alternatively,
one can see the theory as a  conformally improved bosonic string in 2D
target space, with one of the fields self-interacting. Of course this
is just a way to describe our theory and clearly no Lorentz structure
does exist in the ``target space''. The $N-2$ matter fields $\varphi$
can be straightforwardly introduced in the theory by adding to the
previous Lagrangian the term $\partial_+
\varphi_i\partial_-\varphi_i$, $i=3\ldots N$, and to the stress-energy
tensor the term $-(1/2)(\partial_\pm\varphi_i)^2$ (we fix for
convenience in Eq. (\ref{e:grav}) $L=\gamma/8$).

\subsection{The conformal anomaly $c(\gamma)$}
The quantization of the model (\ref{e:liouv_2}) will be performed
using the conventional quantization scheme, which preserves the Weyl
symmetry. In this way we will have a contribution to the anomaly both
from the gravitational and matter field sectors. A consistent
quantization will require a cancellation of the two contributions.

Two different approaches have been proposed in the literature to
realize this cancellation: the 2D quantum gravity {\sl a la} David,
Distler and Kawai \cite{kawai} and the string-inspired dilaton gravity
of Cangemi, Jackiw et al. \cite{jack}.

In Ref. \cite{kawai} the noncritical string theory is considered after
Polyakov's famous paper on the geometry of the bosonic string
\cite{pol}. The string in $d$ dimensions is viewed as a model of $d$
free bosons coupled to 2D quantum gravity. In the functional integral
it is assumed that the measures can all be made independent of the
Liouville mode by a field transformation. The Jacobian so introduced is
supposed to be the exponential of a Liouville action. From these
assumptions it follows the Weyl symmetry of the theory, and this is
sufficient to determine the unknown parameters of the model.  This
approach is consistent for $d\le1$.

In Ref. \cite{jack}, the authors considered a 2D dilaton gravity model
(\ref{e:jack}) with $V(\phi)=\phi$.  In absence of matter fields the
crucial problem of the cancellation of the quantum anomaly has been
solved resorting to an unconventional quantization procedure.  The
action (\ref{e:jack}) is viewed as a theory of two free scalars with
two constraints, where the kinetic energy terms of the two scalars
have opposite sign.  Usually both positive and negative kinetic energy
scalars are quantized imposing positivity of the energy of the quantum
states. This leads to negative norm (ghost) states for the scalar with
wrong sign of the kinetic energy and to the same central charge (+1)
for both scalars. In Ref. \cite{jack} an opposite choice is made,
positivity of the norm is required, leading to negative energy
states. The two scalars contribute with opposite sign to the central
charge and the total conformal anomaly cancels. Of course this method
does not work when matter fields are present.

In this paper we are considering, as in Ref.  \cite{jack}, a dilaton
gravity model, but we will use the conventional quantization procedure
of Ref. \cite{kawai}.  Let us now evaluate the conformal anomaly for
the model (\ref{e:liouv_2}) as a function of $\gamma$. It is standard
lore that a single scalar field $\varphi$ with action $S=(1/8\pi)\int
d^2x \sqrt{|g|}(\partial_\mu\varphi\partial^\mu\varphi
+(2/\gamma)\varphi R)$ has a central charge $c_\pm=1\pm12/\gamma^2$,
depending on the sign of the kinetic energy term (the plus sign holds
in the positive case). $c_+$ is the classical central charge for LFT
$S=(1/4\pi)\int dx^+dx^- (\partial_+\psi\partial_-\psi
+(\mu/4\gamma^2)\exp(\gamma\psi))$. At the quantum level this charge
is shifted \cite{thorn}:
\begin{equation}\label{e:shift}
1+\frac{12}{\gamma^2} \to
1+\frac{12}{\gamma^2}\left(1+\frac{\gamma^2}{2}\right)^2.
\end{equation}
It should be noticed that choosing a dilaton potential $V(\phi)=\alpha
e^{2\gamma\phi}+\beta e^{-2\gamma\phi}$, the central charges for
$\psi$ and $\chi$ (the ghost field) would be shifted by opposite
amounts so that the total anomaly would no longer depend on $\gamma$.
Therefore, there is no way to get  a vanishing anomaly.
A dilaton gravity model with such a potential cannot be, at least 
using our scheme,  quantized.  

Taking into account the shift (\ref{e:shift}), the well-known
contribution from the reparametrization ghosts, and from the $N-2$
matter fields, the total central charge is:
\begin{eqnarray*}
&& c_\psi + c_\chi +(N-2)-26  \\ = &&\left[1
+\frac{12}{\gamma^2}\left(1+\frac{\gamma^2}{2}\right)^2 \right] +
\left[ 1-\frac{12}{\gamma^2} \right] +(N-2)-26  \\
= && 3(\gamma^2-4)+N-2 = c(\gamma).
\end{eqnarray*}
It follows that
\begin{equation}\label{e:anomalia}
c(\gamma)=0 \Rightarrow \gamma=\pm\sqrt{\frac{14-N}{3}}.
\end{equation}
In the following only positive solution will be considered. Since
$\gamma$ has to be real and non-vanishing, we find the upper bound
$N<14$.

\section{Mapping onto a linear dilaton CFT}
We are now in a position to consistently quantize our model. We have
already pointed out that our theory describes a Liouville field
$\psi$, a decoupled free scalar field $\chi$ with the wrong sign
kinetic energy an $N-2$ free scalar matter fields. The only
(self-)interacting field is the Liouville field, whose energy momentum
tensor is the same as a free field with a conformal improvement.  From
Eq. (\ref{e:constr_1}) it is easy to see that also the field $\chi$
has an improvement. The
energy momentum tensor for the quantum theory with $N$ fields is
therefore given by
\begin{equation}\label{e:tensor}
T_{\pm \pm}= -\frac{1}{2}(\partial_\pm X)^2 +v_\mu\partial_\pm^2 X^\mu;
\quad v_\mu = \left( \frac{1}{\gamma},\frac{Q}{2},\mathbf{0}\right),
\end{equation}
where $\mu=0,1,\ldots,N-1$.  $X^0$ is the  field $\chi$, $X^1$ is
the Liouville field $\psi$ and the remaining are the matter
fields. $Q=(2/\gamma + \gamma)$ and $v_\mu$ gives the conformal
improvements. For the $X^\mu$ fields (the ``target space'') we are
using the flat metric $\eta_{\mu\nu} = { \rm diag}(-1,1,\ldots,1)$.  The
energy-momentum tensor (\ref{e:tensor}) can be also derived from the
action:
\begin{equation}\label{e:dilat}
S=-\frac{1}{8\pi}\int d^2 x\sqrt{-g}\left( \partial^a X^\mu\partial_a
X_\mu -2v_\mu X^\mu R\right).
\end{equation}
This action describes a linear dilaton CFT. This is a free field
theory whereas the theory described by Eqs. (\ref{e:liouv_2}) and
(\ref{e:constr_1}) is not. However, both theories have the same
energy-momentum tensor and this property can be used to determine the
spectrum of our model.  Before doing so, we will briefly remind some
basic features of LFT \cite{thorn,bra}, relying mainly on the picture
proposed in Refs. \cite{do,zz} (see also Ref. \cite{tesc}).  In this
approach LFT is viewed as a mild generalization of the standard 2D CFT
structure. The older approach to LFT (see e.g. Ref. \cite{seib}) is
much more involved and cannot be used for our purposes.

\subsection{Liouville Field Theory}
LFT is quantized as a CFT generalizing the celebrated framework of
Ref. \cite{bpz}, henceforth referred to as BPZ. The space of the
states forms a representation of the Virasoro algebra but, in contrast
with the minimal model (BPZ scheme), the set of representations is
continuous. This is due to the noncompactness of the space where the
zero mode of the theory takes values.

A two dimensional CFT is characterized by a correspondence between
fields, i.e. local operators, and states \cite{cft}. The primary
fields $V_\alpha$ acting on the $SL(2,C)$ invariant vacuum $|0\rangle$
generate the highest weight states of the Virasoro algebra. The
descendant fields (states) are defined by the action of the
energy-momentum tensor $T(w)$ (Virasoro modes $L_n$) on the primary
fields (highest weight states). The theory is fully specified by
vacuum expectation values (VEV) of the form:
\begin{equation}\label{e:vacuum}
\langle 0 | \prod_{p=1}^N T(w_p) \prod_{q=1}^M \bar{T}(\bar{w}_q)
\prod_{r=1}^R V_{\alpha_r}(z_r,\bar{z}_r) | 0 \rangle .
\end{equation}
LFT fits into this general scheme but there are some subtleties. 
If the central
charge is given by $c=1+3Q^2$ (in our case $Q=2/\gamma + \gamma$),
then $e^{\alpha\phi}$ are spinless primary fields with conformal
dimension
\begin{equation}\label{e:dim}
\Delta(e^{\alpha\phi}) = \frac{1}{2}\alpha(Q-\alpha).
\end{equation}
The correspondence exists only in the
region
\begin{displaymath}
0 < \Re(\alpha) \le \frac{Q}{2}.
\end{displaymath}
A scalar product can be defined for states $|\alpha \rangle$, defined
as usual by $\lim_{z\to 0} V_\alpha(z) |0\rangle$, if $\alpha$ is
given by
\begin{displaymath}
\alpha = \frac{Q}{2} + iP,
\end{displaymath}
where $P$ is real (we can take $P>0$). As a consequence, using
Eq. (\ref{e:dim}) we find that a state $|\alpha >$ has a conformal
dimension $\Delta(P) = Q^2/8 + P^2/2$. The Hilbert space  the theory
is given by
\begin{equation}\label{e:hilbert}
\bigoplus_{P>0} \text{Vir}_{\Delta(P)}\otimes 
\text{Vir}_{\bar\Delta(P)},
\end{equation}
where $\bigoplus_{P>0}$ is the direct sum over $P>0$ and
$\text{Vir}_{\Delta(P)}$ are irreducible representations of the
Virasoro algebra of highest weight $\Delta(P)$.

From Eq. (\ref{e:hilbert}) it is evident that, paradoxically, the
vacuum $|0\rangle$ does not belong to the Hilbert space.  Nevertheless
the state-operator correspondence and VEV (\ref{e:vacuum}) still make
sense. Owing to the CFT structure of LFT it is in principle
possible to reconstruct all correlation functions starting from the
three-point function, whose exact expression is given in
Ref. \cite{do}. VEV are obtained by suitable analytical continuations
summing over intermediate states \cite{tesc}.

\subsection{The quantization of the linear dilaton CFT}

We have already shown that our dilaton gravity model has the same
energy-momentum tensor of a linear dilaton CFT.  The spectrum of the
theory can be found by quantizing a sort of bosonic string with $N$
bosons $X^\mu$. The energy-momentum tensor and the action being given
by Eqs. (\ref{e:tensor}) and (\ref{e:dilat}).  We can simply follow
the steps that are usual for the critical bosonic string, taking into
account i) the presence of a conformal improvement and ii) the
previously discussed structure of LFT.  Concerning point ii), we take
a ground state (oscillatory vacuum) $|p;0\rangle$ of momentum $p$ with
component $(p^1)^2 > Q^2/8$.

We follow the conventions of Ref. \cite{gsw}.  The worldsheet is
parametrized by $(\tau,\sigma)$, $-\infty < \tau < \infty$, $0\le
\sigma \le \pi$ and we use periodic boundary conditions.  The modes
for right ($\alpha_n^\mu$) and left ($\tilde\alpha_n^\mu$) movers are
independent. The commutation relations are as usual
$[\alpha_m^\mu,\alpha_n^\nu]=[\tilde\alpha_m^\mu,\tilde\alpha_n^\nu]=
m\eta^{\mu\nu}\delta_{m+n}$. The Virasoro operators for $m\ne 0$ are
given by:
\begin{equation}\label{e:vira}
L_m = \frac{1}{2}\sum_q \alpha_{m-q}\cdot\alpha_q -imv\cdot 
\alpha_m.
\end{equation}
Here, and in the following the expressions for the left modes
operators are obtained substituting $\alpha_n \to \tilde\alpha_n$. The
normal ordered expression for $L_0$ is:
\begin{displaymath}
L_0 = \frac{p^2}{8} +\sum_{n=1}^\infty \alpha_{-n}\cdot \alpha_n.
\end{displaymath}
The Hamiltonian is $H = L_0 + \tilde L_0$.  The normal ordering
constant $a$ in the mass-shell conditions $(L_0-a)|\phi\rangle
=(\tilde L_0-a)|\phi\rangle=0$ and the conformal anomaly are
determined in 
a well-known way. From the relations for integer $m$
\begin{displaymath}
\frac{1}{12}[(N+12v^2)m^3-Nm] +2ma+\frac{1}{6}(m-13m^3)=0,
\end{displaymath}
we read
\begin{equation}\label{e:v}
v^2 = \frac{26-N}{12}; \qquad a = \frac{N-2}{24}.
\end{equation}
As expected, the first equation above is the same as
Eq. (\ref{e:anomalia}). From Eq. (\ref{e:v}) it follows immediately that
in the absence of matter fields ($N=2$) the ground state is massless
($a=0$). If matter fields are present ( $3\le N < 14$) the ground
state is tachyonic ($a>0$).

\section{The spectrum of the model}
Let us now construct explicitly the spectrum of our model. We will
first consider the case in which matter fields are present,
$N\ge3$. In this case, we can generate the spectrum and prove the
unitarity of the theory, using a spectrum-generating algebra.  This
can be done adapting to our case the Brower construction \cite{br}, a
version of the covariant formalism of Del Giudice, Di Vecchia and
Fubini \cite{DDF} (see also Ref. \cite{gsw}) used to prove the
no-ghost theorem for the bosonic string.

We consider only the right movers sector (our results can be
immediately extended to the left movers sector).  We want to construct
a spectrum generating set of operators $A^\mu_n$ commuting with the
Virasoro generators $L_m$.  Starting from the right moving solution at
$\sigma=0$
\begin{displaymath}
X^\mu_R(\tau) = \frac{1}{2}x^\mu + \frac{1}{2}\tau p^\mu +
\frac{1}{2}i\sum_{n\ne0} \frac{\alpha_n^\mu}{n}e^{-2in\tau},
\end{displaymath}
we first construct primary fields of conformal dimension $1$ from the
vertex operators $V(k,\tau) = :\lambda\cdot\dot{X}_R\exp(ikX_R):$,
where $\lambda$ is a proper polarization vector. Using
Eq. (\ref{e:vira}) we find:
\begin{eqnarray}\label{e:vertici}
&& [L_m,\,:\lambda\cdot\dot{X}_Re^{ikX_R(\tau)}:] = \nonumber \\
&&e^{2im\tau}\left( -\frac{i}{2}\frac{d}{d\tau} +m(1+\frac{k^2}{2}-
ik\cdot
v)\right):\lambda\cdot\dot{X}_Re^{ikX_R}: \nonumber \\
 && + e^{2im\tau}m^2\left( \frac{k\cdot\lambda}{2}
-iv\cdot\lambda\right):e^{ikX_R(\tau)}:
\end{eqnarray}
Let us now take $k$ lightlike and orthogonal to $v$. This is always
possible since from Eq. (\ref{e:v}) it follows that $v$ is
spacelike. It is convenient to use light-cone coordinates $X^\mu =
(X^+,X^-,X^i)$ with $X_\mu Y^\mu = -X^+ Y^- - X^- Y^+ + X^iY^i$. By
means of a Lorentz rotation, we are free to take
$v_\mu=(0,0,v,0\ldots,0)$, where $v_\mu v^\mu=v^2$. The kinematical
setup is fixed as follows.  The ground state momentum $p_0^\mu$ in
$|p_0;0\rangle$ can be chosen such that $p_0^\mu=(-2,0,\beta,\ldots)$,
$\beta^2= 8a$. The $k$ in $V(k,\tau)$ is $k_n = (0,-2n,\mathbf{0})$
for integer $n$. $k_n^2=0$, $k_n^\mu v_\mu=0$. At the level $n$ the
mass-shell condition is satisfied: $(p+k_n)^2 +8(n-a) =0$.

Let us first construct the operators that generate states describing
excitations of the matter fields ($N\ge 4$).  If $\lambda^i$ is a
vector pointing in the $i$ direction, we can find $N-4$ operators,
\begin{equation}\label{e:norm}
V^i(k_n,\tau) = :\lambda^i\cdot\dot{X}_R e^{ik_nX_R}:, \quad i\ge 4 ,
\end{equation}
which, from Eq. (\ref{e:vertici}), satisfy the commutation relations
\begin{equation}\label{e:buona}
[L_m,V^i(k_n,\tau)] = e^{2im\tau}\left
 ( -\frac{i}{2}\frac{d}{d\tau} +m
 \right)V^i(k_n,\tau).
\end{equation}
The operators of the spectrum-generating algebra $A^\mu_n$ are easily
found to be
\begin{displaymath}
A^\mu_n = \frac{1}{\pi}\int_0^{\pi}d\tau V^\mu(k_n,\tau).
\end{displaymath}
Due to the kinematical setup, the vertex operators $V^\mu$ are
periodic with period $\pi$. It follows that for
$i\ge 4$, $[L_m,A_n^{i}]=0$ .

For $i=3$ a compensating term has to be added to the expression
(\ref{e:norm}), since $\lambda^3\cdot v\ne0$ (see
Eq. (\ref{e:vertici})). The form of this term is well known \cite{br}. If
$\hat{v}$ is the unit vector in the direction $v_\mu$, we have
\begin{displaymath}
V^3(k_n,\tau) = :\hat{v}\cdot\dot{X}_R e^{ik_nX_R}: +
v\frac{d}{d\tau}(\ln k_n\cdot \dot{X}_R)e^{ik_nX_R},
\end{displaymath}
so that $V^3$ satisfies Eq. (\ref{e:buona}) and consequently 
$A^3_n$
commutes with $L_m$. 

Let us now construct the operators $A^+_n$ and $A_n^-$, which generate
states describing excitations in the ${\pm}$ directions. From the
equation $V^+(k_n,\tau)=\dot{X}_R^+\exp(ik_n X_R)$, it follows that
$A^+_n$ is trivial, $A^+_n = (1/\pi)\int_0^\pi
V^+(k_n,\tau)=p^+/2=-\delta_n$.  On the other hand $V^-$ can be defined
as
\begin{displaymath}
V^-(k_n,\tau) = :\dot{X}^-_R e^{ik_nX_R}: -\frac{1}{2}i n
\frac{d}{d\tau}(\ln k_n\cdot \dot{X}_R)e^{ik_nX_R}.
\end{displaymath}

We have completed the construction of the spectrum-generating 
operators $A^\mu_n$. They satisfy the  algebra,
\begin{eqnarray}\label{alg}
&&[A^i_m,A^j_n] = m\delta^{ij}\delta_{n+m} \qquad i,j\ge 3\nonumber
\\
&&[A^-_m,A^j_n] = -nA^i_{n+m} \qquad i > 3  \nonumber\\
&&[A^-_m,A^3_n] = -nA^3_{n+m} -ivn^2\delta_{n+m} \nonumber \\
&&[A^-_m,A^-_n] = (m-n)A^-_{m+n} +2m^3\delta_{n+m}
\end{eqnarray}
Notice that the matter fields act as transverse ``string'' oscillators. 
As usual, instead with
$A^-_n$ it is convenient to work with the operators
\begin{displaymath}
\tilde A^-_n = A^-_n
-\frac{1}{2}\sum_{p=-\infty}^{\infty}\sum_{i=1}^{N-2}:A^i_{n-p}A^i_{p}:
+\delta_n\frac{\beta^2}{8}.
\end{displaymath}
The operators $\tilde A^-_n$ commute with $A^i_m$, by definition they 
annihilate the oscillator vacuum,
\begin{equation}\label{m}
\tilde A^-_0|p_0;0\rangle =0.
\end{equation}
Furthermore $\tilde A_n^-$ obey a Virasoro
algebra:
\begin{displaymath}
[\tilde A^-_m, \tilde A^-_n] = (m-n)\tilde A^-_{m+n}.
\end{displaymath}
We have successfully completed our task of finding the spectrum of 
the model we are considering. From the algebra (\ref{alg}) it follows 
that for $N\ge3$ the spectrum is that of a bosonic string oscillating 
in a target space with  $N-2$ transverse direction.
The gravitational sector is decoupled 
from the matter field sector.  Eq. (\ref{m}) implies  that the 
states of this sector have zero norm, as expected for   
pure gauge states.

Until now we have considered only the case when matter field are
present ($N\ge3$).  If matter fields are absent ($N=2$), we cannot use
the previously explained construction.  However, in this case all the
excitation are pure gauge and it is evident that the Hilbert space
consists of the ground state whereas all the excited states have zero
norm.  This conclusion is consistent with the results for pure
(Jackiw-Teitelboim) dilaton gravity obtained in Ref. \cite{jack}, using
a different approach.

\section{Conclusions}
In this paper we have consistently quantized a model of matter-coupled
dilaton gravity in two dimensions with exponential dilaton potential.
A vanishing conformal anomaly has been achieved by tuning a parameter
in the dilaton potential. The quantization has been performed by
mapping the theory first onto a field theory with a Liouville interaction
and then onto a linear dilaton CFT.  The spectrum has been determined
in a a straightforward way, analogous to that used for the bosonic
string in critical dimensions.

We have found that the ground state is tachyonic (or massless in
absence of matter). The spectrum has two decoupled sectors: the
gravitational sector made of pure gauge, zero norm states and the
matter field sector describing transverse physical excitations.  This
result confirms previous results \cite{jack} about quantization of
pure dilaton gravity models. 

The theory has a free field spectrum but it is not trivial, since as
far as correlations are concerned, it has at least the same complexity
as LFT. With our approach, establishing the equivalence with a sort of
critical string, we have  succeeded in what seems difficult
using other methods: the quantization of matter-coupled dilaton
gravity.

At first sight for pure 2D dilaton gravity 
our results seem to contradict the results of Ref. 
\cite{klw}. In that paper the authors have shown, using path integral 
quantization, that for all 2D dilaton (pure) gravity models the quantum 
effective action coincides with the classical one. 
However, it is not difficult to realize that the results of our paper 
and those of Ref. \cite{klw} are not necessarily in contradiction.
Using our approach we can make statements only about  the
spectrum not about the correlation function of the model, so that
we cannot say anything about 
the quantum effective action. 
Moreover, as  long as the spectrum is concerned, for pure gravity
our result is the same as that of Ref. \cite{klw}: the theory is 
locally trivial also at the quantum level, all the local excitations  
are pure gauge.

As a final point, we observe that the rich structure exhibited by the
semiclassical analysis of 2D dilaton gravity has disappeared.
The requirement of vanishing
anomaly, i.e. the criticality of the theory, washes out the
semiclassical structures. The gap between the semiclassical and the
quantum theory   has still to be filled. We believe this
can be done only going off-criticality. In this way the theory would be
equivalent to a noncritical string and needless to say would present
severe difficulties.

\begin{acknowledgments}
Discussions with R.~Jackiw and M.~Cavagli\`a are warmly
acknowledged. P.C. is grateful to the Center for Theoretical Physics,
MIT, for hospitality during part of this work.
\end{acknowledgments}

\end{document}